\documentclass[12pt]{article}
\usepackage[utf8]{inputenc}
\usepackage[T1]{fontenc}
\usepackage{geometry}
\usepackage{setspace}
\usepackage{amsmath, amssymb}

\usepackage{graphicx}
\usepackage{caption}
\usepackage{booktabs}
\usepackage{longtable}
\usepackage{pdflscape}
\usepackage{enumitem}

\usepackage{xcolor}
\usepackage{color}
\usepackage{listings}
\usepackage{titlesec}
\usepackage{fancyhdr}
\usepackage{hyperref}

\usepackage{algorithm}
\usepackage{algorithmic}
\usepackage[most]{tcolorbox}

\usepackage{cite}
\usepackage{natbib}
\lstdefinestyle{custompython}{
  language=Python,
  basicstyle=\ttfamily\small,
  keywordstyle=\color{blue},
  commentstyle=\color{gray},
  stringstyle=\color{orange},
  numbers=left,
  numberstyle=\tiny\color{gray},
  stepnumber=1,
  numbersep=8pt,
  breaklines=true,
  frame=single,
  captionpos=b,
  morekeywords={INPUT, PROCESS, FOR, AGGREGATE}
}

\title{Canary in the Mine: An LLM Augmented Survey of Disciplinary Complaints to the Ordre des ingénieurs du Québec (OIQ)}

\author{
  Tammy Mackenzie\thanks{Corresponding author: tammy@theaulafellowship.org}, Varsha Kesavan, Thomas Mekhael, Animesh Paul, \\Branislav Radeljic, Sara Kodeiri, Sreyoshi Bhaduri\\}

\date{}

\begin{document}

\maketitle

\begin{abstract}

This study uses LLMs to investigate disciplinary incidents involving engineers in Quebec, shedding light on critical gaps in engineering education. Through a comprehensive review of the disciplinary register of the Ordre des ingénieurs du Québec (OIQ)'s disciplinary register for 2010 to 2024, researchers from engineering education and human resources management in technological development laboratories conducted a thematic analysis of reported incidents to identify patterns, trends, and areas for improvement. The analysis aims to uncover the most common types of disciplinary incidents, underlying causes, and implications for the field in how engineering education addresses (or fails to address) these issues. Our findings identify recurring themes, analyze root causes, and offer recommendations for engineering educators and students to mitigate similar incidents. This research has implications for informing curriculum development, professional development, and performance evaluation, ultimately fostering a culture of professionalism and ethical responsibility in engineering. By providing empirical evidence of disciplinary incidents and their causes, this study contributes to evidence-based practices for engineering education and professional development, enhancing the engineering education community's understanding of professionalism and ethics.
\end{abstract}

\section{Introduction}
The engineering profession is built on a foundation of trust, integrity, and ethical conduct. However, like any profession, engineering is not immune to instances of misconduct, negligence, and unethical behavior. The Ordre des ingénieurs du Québec (OIQ)'s disciplinary register provides a unique window into the types of complaints filed against engineers in Quebec. Much like a canary in a coal mine, these complaints act as early warning signals—symptoms of deeper, systemic issues within the profession that may otherwise go unnoticed. By analyzing these complaints, we can gain insight into the challenges facing the engineering profession and identify areas for improvement in engineering education. By analyzing complaints against engineers, we gain valuable insights into the profession’s challenges and can highlight key areas for improvement in engineering education. One key takeaway is the need for students to be exposed to real-world examples of what constitutes misconduct, rather than relying solely on abstract contexts, a charge traditional engineering ethics is often guilty of.\\

This study uses large language models (LLMs) to analyze 387 disciplinary cases from the Ordre des ingénieurs du Québec (OIQ) (2010–2024). We identify recurring misconduct themes and assess how such violations can inform engineering ethics education. Results suggest the need for deeper integration of real-world case studies into engineering curricula to prevent similar infractions. Specifically, our analysis reveals a range of offenses. We provide an in-depth examination of the article codes under which disciplinary decisions are rendered, shedding light on the most common types of complaints and the corresponding penalties.\\

The findings of this study have significant implications for engineering education, highlighting the need for a pedagogical framework that addresses these disciplinary challenges. By integrating ethics, professionalism, and accountability into engineering curricula, educators can help foster a culture of integrity and responsibility within the profession. This research is of potential use for engineering educators, managers of engineers, engineering students, and deontology scholars seeking to promote ethical conduct and professionalism in engineering practice. The contributions of our paper are trifold:
\begin{enumerate}
    \item A comprehensive analysis leveraging Large Language Models for human-in-loop thematic analysis of complaints filed against engineers in Quebec, providing quick insight from the real world, into the types of violations that occur in the profession.
    \item Implications for engineering educators that addresses disciplinary challenges in engineering education, highlighting the need for ethics, professionalism, and accountability in engineering curricula.
    \item Implications for professional orders wishing to decrease misconduct by engaging with the finer details of the training process. 
\end{enumerate}

\section{Background}

Engineers face various ethical dilemmas in the workplace, ranging from infrequent to frequent occurrences \citep{bielefeldt2016changes}. These dilemmas often involve conflicts between profit motives and public good, leading some engineers to change careers \citep{bielefeldt2016changes}. Common ethical issues include illegal waste dumping and data manipulation \citep{chalk1989drawing}. Research suggests a concerning link between academic dishonesty in engineering education and unethical behavior in professional settings. Studies have found that engineering students are among the most likely to engage in academic cheating \citep{carpenter2006engineering, carpenter2006implications}. This behavior appears to correlate with unethical conduct in the workplace, as demonstrated by surveys exploring decision-making patterns in both academic and professional contexts \citep{harding2003relationship}. These findings highlight the need for interventions to address unprofessional behavior. Researchers have identified various approaches, with most interventions targeting individuals rather than organizations and focusing on increasing awareness \citep{tricco2018prevention}. Sexual harassment is also often the most frequently targeted behavior for change. While several promising interventions exist, the majority consist of single-component, in-person education sessions, with fewer studies addressing institutional culture or behavior change directly \citep{tricco2018prevention}.\\

Studies also indicate that engineering students self-report high frequencies of cheating in college \citep{harding2003relationship}. This behavior correlates with academic dishonesty in high school and other deviant behaviors like petty theft and lying \citep{carpenter2006implications}. Multiple investigations have explored the relationship between academic cheating and workplace ethics, finding similarities in decision-making processes across academic and professional contexts. Frequent cheaters in high school were more likely to violate workplace policies \citep{harding2003relationship}. These findings raise significant concerns for engineering educators, corporations, and society, highlighting the need to address academic integrity to promote ethical professional conduct \citep{carpenter2006implications}. The research emphasizes the importance of understanding students' perceptions of cheating and developing strategies to increase academic integrity in engineering education. Professional licensing boards play a crucial role in regulating the engineering profession, establishing ethical standards, and taking disciplinary actions against violations \citep{marsico2009professional}. Efforts to promote ethical behavior among engineering students include incorporating ethics education into curricula and introducing initiatives like the Engineers' Affirmation, similar to the Hippocratic Oath \citep{carter2011making}. These approaches aim to influence students' ethical decision-making and reduce unprofessional conduct throughout their careers. Implementing such measures in engineering education may help cultivate a stronger ethical foundation for future professionals and mitigate unethical behavior in both academic and workplace settings.Disciplinary reports in engineering orders thus provide valuable insights into the professional misconduct of engineers. The Disciplinary Council of the Ordre des ingénieurs du Québec (OIQ) is an autonomous and independent administrative tribunal entrusted with a critical mission: protecting the public in Quebec, Canada. As a key component of Quebec's engineering regulatory framework, the Disciplinary Council plays a vital role in ensuring that engineers in the province adhere to the highest standards of professional conduct and ethics. 

\section{Methods}

\subsection{Introduction and Data Curation}
The Disciplinary Council of the Ordre des ing\'enieurs du Qu\'ebec (OIQ) is responsible for adjudicating complaints concerning ethical violations in the professional conduct of its members. These complaints may relate to the Code des professions, the Engineers Act, or the Code of Ethics of Engineers, the latter being a regulation under the former two acts. Since June 1, 2001, the Council's decisions have been made publicly available via the SOQUIJ website~\citep{soquij_portal}. For this study, we collected disciplinary decisions issued between 2010 and 2024 from the SOQUIJ database. Our focus is on cases involving violations of the Code of Ethics of Engineers. We employ a large language model (LLM)-augmented thematic analysis (e.g., \citep{bhaduri2024reconciling, kapoor2024qualitative, bedemariam2025potential}) to identify recurring patterns in ethical breaches. Specifically, we extract the unique violated articles from each judgment and categorize them according to thematic groupings defined in the legal text~\citep{quebec_ethics_code}. These violations are then analyzed across three time periods: 2010--2014, 2015--2019, and 2020--2024.

\subsection{Summarization Using LLMs}
We used Google's Gemini 2.0 Flash model~\citep{team2023gemini}, accessed via LangChain~\citep{langchain}, a framework for integrating LLMs in processing pipelines. LangChain utilities include prompt templates, document loaders, and integrations with cloud platforms like Google Vertex AI. The disciplinary judgment documents were downloaded and converted into plain text using LangChain's \texttt{UnstructuredWordDocumentLoader}. Texts were summarized using the "stuffing" method in a zero-shot setting with the following prompt:

\begin{quote}
\textit{You are a helpful assistant that summarizes text. Summarize the following text identifying all the charges and counts: \{text\}. Once summarized, include a summary of the circumstances, as well as identify all charges the accused is ultimately found guilty of.}
\end{quote}

Our large language model (LLM)-augmented workflow for extracting violated legal articles from disciplinary judgments involves the following structured sequence:\\

\begin{tcolorbox}[title=LLM-Augmented Article Extraction Workflow, colback=gray!5, colframe=black!40, coltitle=black, fonttitle=\bfseries]
\begin{algorithmic}[1]
\REQUIRE Disciplinary records in \texttt{.docx} or \texttt{.doc} format (387 cases from SOQUIJ~\citep{soquij_portal})
\FOR{each judgment document}
    \STATE Convert to plain text using \texttt{UnstructuredWordDocumentLoader} (LangChain)
    \STATE Summarize document using Gemini 2.0 Flash via \texttt{load\_summarize\_chain} in LangChain\\
           \hspace{1.5em}(using zero-shot prompting and the ``stuffing'' method)
    \STATE Apply a one-shot prompt to extract guilty charge descriptions and article references
    \STATE Use Python \texttt{regex} to extract legal article numbers (e.g., \texttt{"2.01"}, \texttt{"3.01.01"})
    \STATE Record only unique articles per judgment
\ENDFOR
\STATE Aggregate all unique article references
\STATE Group articles by time period: \texttt{2010--2014}, \texttt{2015--2019}, \texttt{2020--2024}
\STATE Compute frequency distribution for each article across time periods
\STATE Output results as a histogram by theme and time window
\end{algorithmic}
\end{tcolorbox}

\subsection{Guilty Charge and Article Extraction}
We further prompted Gemini using one-shot prompting to extract guilty charges and their corresponding legal articles. A sample input and expected output were specified in the prompt to guide formatting. Using Python's regex library \citep{van1995python}, we extracted article references such as ``2.01'' or ``3.02.03'' from the model's output. To ensure consistency, each article was counted only once per judgment regardless of multiple appearances. This approach helps mitigate over-representation of any article due to repetition in a single case. The extracted articles were then mapped to thematic categories as outlined in the Code of Ethics of Engineers~\citep{quebec_ethics_code}. This allowed us to aggregate the violations across time periods and perform structured comparisons. We adopted a unique-per-judgment counting method—counting each article only once per disciplinary case, regardless of how many times it was cited within that case. This approach provides a clearer view of how widespread specific ethical breaches are across the profession, helping to identify which articles are most frequently violated in a diverse set of contexts. By contrast, a total-count method would disproportionately weight judgments with multiple charges under the same article, potentially skewing the results toward a few severe cases. While this unique-per-judgment approach is more appropriate for detecting systemic patterns in professional misconduct, we acknowledge that the total-count perspective could be valuable in future analyses aimed at understanding the cumulative burden or enforcement intensity associated with specific regulations.


\subsection{Evaluation of Extraction Accuracy}
Our task involves extracting the set of uniquely violated legal articles from disciplinary judgment documents. To evaluate this, we conducted a task-based evaluation, comparing the workflow's extracted outputs (from generated summaries) against manually annotated sets derived directly from the original judgments. We assumed an expected accuracy of approximately 90\%, based on Gemini 2.0 Flash's strong factual consistency observed during preliminary tests on small case sets. As a result, a sample size of 77 was estimated to be sufficient to obtain an accuracy estimate within ±6\% with 95\% confidence. To ensure year-wise representation across the 2010–2024 period, we applied a proportional sampling strategy where 77 files were sampled from the 387 disciplinary records such that each year contributed in proportion to its presence in the full dataset. Each sampled file was manually reviewed, and the true set of violated legal articles was annotated to serve as reference labels.

\subsection{Evaluation Metrics and Results}

To assess the performance of our article extraction pipeline, we conducted a task-based evaluation using four standard metrics from information retrieval and natural language processing: \textbf{Jaccard accuracy}, \textbf{precision}, \textbf{recall}, and \textbf{F1-score}. These metrics were chosen to quantify the degree of alignment between the set of legal articles extracted by the LLM and the ground truth article sets derived through manual annotation.

\begin{itemize}
  \item \textbf{Jaccard Accuracy} measures the similarity between the predicted set of violated articles ($A_{\text{pred}}$) and the reference set ($A_{\text{true}}$). It is defined as:
  \[
  \text{Jaccard Accuracy} = \frac{|A_{\text{pred}} \cap A_{\text{true}}|}{|A_{\text{pred}} \cup A_{\text{true}}|}
  \]
  This metric is particularly suitable for evaluating set-based tasks, as it penalizes both false positives and false negatives.

  \item \textbf{Precision} is the proportion of predicted article references that are correct. It is defined as:
  \[
  \text{Precision} = \frac{\text{True Positives}}{\text{True Positives} + \text{False Positives}}
  \]

  \item \textbf{Recall} measures the proportion of actual article references that were correctly predicted, calculated as:
  \[
  \text{Recall} = \frac{\text{True Positives}}{\text{True Positives} + \text{False Negatives}}
  \]

  \item \textbf{F1-score} is the harmonic mean of precision and recall, providing a single metric that balances both:
  \[
  \text{F1-score} = 2 \cdot \frac{\text{Precision} \cdot \text{Recall}}{\text{Precision} + \text{Recall}}
  \]
\end{itemize}

Using a stratified sample of 77 disciplinary decisions spanning 2010 to 2024, we manually annotated the true set of violated articles for each judgment and compared them to the LLM-extracted sets. The results were as follows: Jaccard Accuracy: 91.88\%, Precision: 91.88\%, Recall: 100.00\%, and F1-score: 92.89\%. These results demonstrate that the LLM-based extraction pipeline performs with high fidelity. The perfect recall indicates that no relevant article references were missed (i.e., no false negatives), while the high precision reflects relatively few false positives.  

False positives stemmed mainly from:
\begin{enumerate}[label=(\arabic*)]
  \item Ambiguity in articles under suspended proceedings
  \item Misinterpretation of restated charges during sentencing
\end{enumerate}

Despite these edge cases, the evaluation supports the reliability of the LLM workflow for structured extraction of violated legal articles from disciplinary decisions. Manual annotations included only clearly adjudicated guilty charges. These results validate the LLM pipeline’s high fidelity for ethical violation extraction.

\section{Results and Discussion}
Our thematic analysis, as tabulated in this section, of the Ordre des ingénieurs du Québec (OIQ)'s disciplinary register from 2010 to 2024 reflects a concerning landscape of professional misconduct, spanning issues such as collusion, corruption, professional negligence, bribery, fraud, and other ethical breaches. This analysis contextualizes these infractions within historical developments in the construction and engineering industries, highlighting systemic challenges and the limitations in enforcement mechanisms.

\subsection{Overview of Complaints Filed Against Engineers in Quebec}


Our analysis of 668 violations for which engineers were found guilty between 2010 and 2024 reveals persistent trends across categories of ethical breaches. The most common set of violations involve breaches of public and client trust, particularly around integrity, diligence, and disinterestedness. The most frequently cited article across all time periods was \textbf{3.02.08}—\textit{engaging in practices that contravene the dignity or integrity of the profession}—with 73 total violations (11 in 2020–2024, 55 in 2015–2019, and 7 in 2010–2014). Closely following was \textbf{2.04}, which concerns the duty to act with honesty and transparency toward the public, cited 66 times overall. These patterns point to systemic issues related to professional integrity and public accountability. \\

Violations of general integrity obligations toward customers (e.g., Articles 3.02.01, 3.02.04) were also prominent. Article \textbf{3.02.01}, which addresses respect and honesty in dealings with clients, appeared in 93 instances, peaking during 2015–2019. Article \textbf{3.02.04}, concerning misrepresentation or distortion of facts, remained consistently cited across all periods (38 in 2020–2024, 11 in 2015–2019, and 11 in 2010–2014).\\

\begin{landscape}

\begin{longtable}{p{3cm}p{9cm}rrr}
\caption[]{Violations tabulated across themes)} \\
\toprule
\textbf{Article} & \textbf{Code - Theme} & \textbf{2020--2024} & \textbf{2015--2019} & \textbf{2010--2014} \\
\midrule
\endfirsthead
\caption[]{(continued) Violations: Public and Customer Duties (General, Integrity, Diligence)} \\
\toprule
\textbf{Article} & \textbf{Code - Theme} & \textbf{2020--2024} & \textbf{2015--2019} & \textbf{2010--2014} \\
\midrule
\endhead
\midrule
\multicolumn{5}{r}{{Continued on next page}} \\
\midrule
\endfoot
\bottomrule
\endlastfoot
2.01 & Duties and Obligations towards the public & 25 & 16 & 10 \\
2.01(Regulation) & Duties and Obligations towards the public & 1 & 0 & 0 \\
2.01 a) & Duties and Obligations towards the public & 1 & 1 & 1 \\
2.01 b) & Duties and Obligations towards the public & 2 & 1 & 1 \\
2.01 c) & Duties and Obligations towards the public & 3 & 1 & 1 \\
2.03 & Duties and Obligations towards the public & 1 & 0 & 0 \\
2.04 & Duties and Obligations towards the public & 41 & 10 & 15 \\
3.01.01 & Duties and Obligations towards the customer - General provisions & 12 & 13 & 4 \\
3.01.02 & Duties and Obligations towards the customer - General provisions & 0 & 1 & 0 \\
3.01.03 & Duties and Obligations towards the customer - General provisions & 0 & 1 & 0 \\
3.01.04 & Duties and Obligations towards the customer - General provisions & 0 & 0 & 1 \\
3.02.01 & Duties and Obligations towards the customer - Integrity & 26 & 49 & 18 \\
3.02.02 & Duties and Obligations towards the customer - Integrity & 1 & 2 & 1 \\
3.02.03 & Duties and Obligations towards the customer - Integrity & 0 & 3 & 0 \\
3.02.04 & Duties and Obligations towards the customer - Integrity & 38 & 11 & 11 \\
3.02.05 & Duties and Obligations towards the customer - Integrity & 1 & 0 & 0 \\
3.02.06 & Duties and Obligations towards the customer - Integrity & 0 & 0 & 1 \\
3.02.08 & Duties and Obligations towards the customer - Integrity & 11 & 55 & 7 \\
3.02.09 & Duties and Obligations towards the customer - Integrity & 4 & 12 & 0 \\
3.02.10 & Duties and Obligations towards the customer - Integrity & 0 & 7 & 2 \\
3.03.01 & Duties and Obligations towards the customer - Availability and Diligence & 13 & 1 & 3 \\
3.03.02 & Duties and Obligations towards the customer - Availability and Diligence & 4 & 1 & 1 \\
3.03.03 & Duties and Obligations towards the customer - Availability and Diligence & 8 & 1 & 0 \\
3.03.04 & Duties and Obligations towards the customer - Availability and Diligence & 2 & 0 & 0 \\
3.03.05 & Duties and Obligations towards the customer - Availability and Diligence & 3 & 0 & 0 \\
3.04.01 & Duties and Obligations towards the customer - Seal and signature & 13 & 8 & 10 \\
3.04.02 & Duties and Obligations towards the customer - Seal and signature & 6 & 3 & 1 \\
3.05.01 & Duties and Obligations towards the customer - Independence and disinterestedness & 2 & 2 & 2 \\
3.05.02 & Duties and Obligations towards the customer - Independence and disinterestedness & 1 & 3 & 3 \\
3.05.03 & Duties and Obligations towards the customer - Independence and disinterestedness & 6 & 39 & 9 \\
3.05.04 & Duties and Obligations towards the customer - Independence and disinterestedness & 0 & 0 & 1 \\
3.06.01 & Duties and Obligations towards the customer - Professional secrecy & 2 & 2 & 1 \\
3.06.03 & Duties and Obligations towards the customer - Professional secrecy & 1 & 2 & 1 \\
3.06.04 & Duties and Obligations towards the customer - Professional secrecy & 0 & 0 & 1 \\
3.07.01 & Duties and Obligations towards the customer - Accessibility and rectification of files and delivery of documents & 1 & 0 & 0 \\
3.07.06 & Duties and Obligations towards the customer - Accessibility and rectification of files and delivery of documents & 1 & 0 & 2 \\
3.08.01 & Duties and Obligations towards the customer - Fixing and payment of fees & 2 & 0 & 0 \\
3.08.02 & Duties and Obligations towards the customer - Fixing and payment of fees & 1 & 0 & 0 \\
3.08.03 & Duties and Obligations towards the customer - Fixing and payment of fees & 10 & 2 & 2 \\
3.08.04 & Duties and Obligations towards the customer - Fixing and payment of fees & 4 & 1 & 0 \\
4.01.01 a) & Duties and Obligations towards the Profession - \textit{Derogatory acts} & 3 & 3 & 10 \\
4.01.01 c) & Duties and Obligations towards the Profession - \textit{Derogatory acts} & 2 & 1 & 0 \\
4.01.01 f) & Duties and Obligations towards the Profession - \textit{Derogatory acts} & 0 & 0 & 1 \\
4.01.01 g) & Duties and Obligations towards the Profession - \textit{Derogatory acts} & 1 & 0 & 1 \\
4.02.02 & Duties and Obligations towards the Profession - \textit{Relationship with the Order and the confreres} & 7 & 5 & 4 \\
4.02.03 & Duties and Obligations towards the Profession - \textit{Relationship with the Order and the confreres} & 1 & 2 & 3 \\
4.02.03 a) & Duties and Obligations towards the Profession - \textit{Relationship with the Order and the confreres} & 0 & 1 & 1 \\
4.02.03 c) & Duties and Obligations towards the Profession - \textit{Relationship with the Order and the confreres} & 0 & 9 & 0 \\
4.02.04 & Duties and Obligations towards the Profession - \textit{Relationship with the Order and the confreres} & 1 & 0 & 1 \\
4.02.05 & Duties and Obligations towards the Profession - \textit{Relationship with the Order and the confreres} & 1 & 0 & 0 \\
5.01.01 & Obligations relating to the name of engineering companies - \textit{Advertising and representation} & 1 & 0 & 2 \\
5.01.02 & Obligations relating to the name of engineering companies - \textit{Advertising and representation} & 1 & 0 & 0 \\
5.02.03 & Obligations relating to the name of engineering companies - \textit{Name of engineering companies} & 0 & 0 & 1 \\

\end{longtable}

\end{landscape}

We also observe a meaningful cluster of violations around procedural diligence and technical responsibility. Articles such as \textbf{3.03.01--3.03.05} (relating to diligence in practice and availability to clients) and \textbf{3.04.01--3.04.02} (relating to the misuse of seals and professional signatures) appear a combined 59 times. This suggests lapses in administrative responsibility and basic procedural compliance.Violations connected to independence and conflict of interest—particularly \textbf{3.05.03}—are striking. This article alone accounts for 54 citations, with a significant spike (39) in the 2015–2019 period, suggesting perhaps delayed enforcement or increased scrutiny post-Charbonneau Commission.\\

Ethical failures related to professional secrecy (\textbf{3.06.x}), accessibility of documents (\textbf{3.07.x}), and fee transparency (\textbf{3.08.x}) were less prevalent, yet notable in their consistency across time. Violations of fee-related provisions were cited 19 times, with a concentration in the most recent years, possibly reflecting increased client pushback or regulatory enforcement. Finally, derogatory acts under \textbf{4.01.x} and failures to cooperate with the Order under \textbf{4.02.x} reveal broader concerns about intra-professional conduct and the engineer’s relationship with oversight institutions. Articles under 4.01 and 4.02 were cited 49 times in total, including cases of misrepresentation, non-collaboration, and disrespect toward peers and the regulatory body.\\

In sum, the longitudinal data suggests that while the specific types of violations may fluctuate, systemic issues of professional integrity, independence, and procedural diligence remain deeply rooted in the profession. These patterns reinforce the need for engineering education to focus not only on technical proficiency, but also on ethical responsibility, transparency, and regulatory literacy.

\subsection{Implications for Engineering Educators}

Long before a bridge collapses or a contract is corrupted, warning signs emerge—not in headlines, but in quiet rulings by professional orders. These disciplinary cases tell a deeper story. They are the canary in the mines that need to make their way into how we teach and learn engineering. Engineering ethics curricula in at least some institutions already implement the case-based approach, but curriculum approaches focus more on compliance issues than on the ethical questions raised by these practices. This approach is not sufficient, and the place of ethical reflection should be greater in training.\\

To be considered are aspects such as the number of courses and hours of training specifically dedicated to ethics, transversal approach, place of ethics in the rest of the training. One of the central challenges in engineering education is ensuring that students cultivate a strong ethical foundation alongside their technical competencies. Our findings highlight a troubling prevalence of integrity violations, professional negligence, and ethical misconduct, suggesting that traditional approaches to ethics education—often confined to theoretical discussions—are insufficient. While data analysis is necessarily constrained to recorded infractions, it is crucial not to disproportionately emphasize one category of misconduct over another. The cases we examined only reflect reported and adjudicated violations, meaning that numerous infractions may go undetected due to investigative and enforcement limitations. As engineering educators, we argue that a more comprehensive approach is required—one that actively strengthens ethical standards, fosters transparency, and prioritizes rigorous oversight in both academic and professional settings.\\

A curriculum that integrates real-world case studies  can serve as a powerful tool for instilling professional accountability in students. Engineering education should embed discussions on multi-disciplinarity, integrity, and professional responsibility throughout core technical coursework \citep{bhaduri2024multi, carrico2023preparing, mackenzie2024beyond, khan2024path}. Additionally, ethics training should not merely frame misconduct as an individual failing but should contextualize it within the broader consequences for public trust, professional credibility, and systemic integrity. Without meaningful reforms, engineering programs risk producing graduates who are technically proficient yet ill-equipped to navigate the ethical dilemmas they will inevitably face in their careers. Addressing these challenges requires a paradigm shift in engineering education—one that acknowledges ethics as an integral component of technical competence rather than an ancillary concern.\\

To illustrate these points, it is important to understand the difference between compliance and ethics, which are two essential but distinct concepts, especially in contexts like engineering or other professional fields. Compliance is part of ethics, but ethics is broader than mere compliance. Compliance refers to adhering to rules, laws, standards, or procedures established by an external authority. It is often a minimal requirement to avoid sanctions or legal consequences. For example, an engineering project must comply with safety standards established by regulators to be approved. In contrast, ethics concerns the moral principles and values that guide a person's or organization's actions and decisions. It goes beyond mere compliance with rules to consider what is morally right or good. In the field of engineering, for example, an ethical decision might involve prioritizing public safety even if it means exceeding minimal compliance requirements. In summary, compliance focuses on adherence to external rules, while ethics encompasses the internal principles of right and wrong that influence choices and actions beyond legal obligations \citep{daoust2024ethique}.\\

In the field of engineering education, compliance is present throughout the curriculum, as it ensures that future engineers learn to adhere to the standards and regulations that govern their practice. By integrating compliance into training, it ensures that engineers will know and respect established rules, which reduces the risk of illegal or harmful behaviors in the future. For example, an engineer trained to follow safety standards or environmental protocols is less likely to commit violations that could endanger human lives or cause environmental damage. This might, at first glance, seem sufficient to prevent future violations, as it seems logical that if engineers know and follow the rules, they will act accordingly. However, this approach has an important limitation: compliance covers only explicit and often minimal rules. It does not address more complex situations where rules are vague, ambiguous, or non-existent, or when there are conflicts between different standards (for example, between the obligation to follow a technical standard and the imperative to protect the environment).\\

This is where ethics training becomes essential. Unlike compliance, which focuses only on following rules, ethics helps engineers develop deeper reflection on their choices. It teaches them to evaluate the social, environmental, and human consequences of their actions, even in situations where there is no explicit regulation to follow. For example, an engineer might face an ethical dilemma where safety standards are met, but the environmental impact of a project is neglected. Ethics allows the engineer to consider these other dimensions and make informed choices that go beyond legal requirements. Thus, while compliance is necessary to ensure a minimum level of regulation, ethics training is what enables engineers to navigate complex situations and make morally responsible decisions. \\

The data collected seems to support this. The highest rates of infractions are found in practices that are easily demonstrable, where the rules are clear. Conversely, the rates are lower when it comes to infractions that are harder to demonstrate. These figures should not lead to the conclusion that it is more important to focus on negligence than on corruption. Indeed, these figures do not inform us about the severity of the infractions, only their number. Thus, while it may be true that the emphasis on compliance can help in cases of certain infractions such as negligence, it is rather ethics training that can help prevent other types of infractions such as exploitation, corruption, or fraud.\\

In this regard, this in one of the reasons as to why ethics courses are now mandatory for obtaining an engineering degree in Canada. According to the 2019 BCAPG requirements, the objective of this training is to enable students to master two essential competencies: first, professionalism, which refers to understanding the roles and responsibilities of engineers in society, particularly in terms of public protection and general interest; second, deontology and equity, which refer to understanding and respecting the principles of professional ethics and responsibility \citep{engineerscanada_accreditation}. However, the time allocated to these courses in engineering training varies from one university to another. Generally, ethics training is concentrated in a single course in the students' curriculum, varying from 1 credit (45 hours) to 3 credits (135 hours) over a single session. These courses are often seen as marginal to the rest of the more technical engineering training. As Bégin points out \citep{begin2006professionnalisme}, integrating ethics throughout the training program is crucial. The idea is to help students understand that ethics and professionalism are integral parts of their professional identity. Too often, however, ethics training is perceived as an external element to the discipline, which can reduce its importance in the eyes of students. However, for this training to have a real impact, it must be seen as a tool for building the professional identity of engineers. \\

Raising awareness among future engineers about the specific ethical issues of their field is an effective way to prevent them from reproducing problematic behaviors, customs, or practices that exist in the professional environment. It helps them not to tolerate practices that could undermine the integrity of their work or lead to harmful consequences for society. To this end, concrete proposals can be made, such as increasing the number of ethics training hours in the curriculum, giving greater importance to ethics in internship reports and final projects, etc.

\subsection{Implications for Engineering Orders}
These findings can inform the professional order of a given region on participating in and advising on potential paths for training adjustments at regional training institutions like universities and trade schools. This is an interactive process of creation of a social culture for engineers, such as suggested by whole-education and other social systemic approaches, for example Scandinavian Institutionalism’s concept of translation, wherein processes of social change are iterative and mutually constructed between different organisations in the field \citep{olsen1997institutional}. Further research would examine the specific context of the Quebec law for professional orders, which is particular to Quebec within north America, and can provide novel legislative acts to support the integration of training interests with professional deontology, within Quebec but also as an inspiration for potential changes elsewhere. 

\subsection{Actionable Recommendations}

Building on our findings, we propose a set of actionable recommendations for both engineering educators and professional orders to strengthen ethical standards within the profession.
\subsubsection{For Engineering Educators:}
\begin{itemize}
    \item Expand ethics instruction: Increase the number of hours dedicated to ethics education and integrate ethical reasoning consistently across technical coursework, internships, and capstone projects to reinforce its relevance throughout the curriculum.
    \item Enhance pedagogical methods: Move beyond theoretical discussions by incorporating real-world disciplinary cases, role-playing exercises, and ethical simulations. These applied strategies will foster deeper critical thinking and prepare students to navigate complex ethical dilemmas.
    \item Foster ethical identity: Position ethics as a core component of professional identity formation. Ethics should be presented not as a peripheral topic but as a foundational element of what it means to be an engineer, cultivating a sense of responsibility and accountability from the outset of training.
\end{itemize}

\subsubsection{For Professional Orders:}
\begin{itemize}
    \item Advocate for curriculum reform: Collaborate with educational institutions to encourage the integration of ethics throughout the curriculum, using empirical data on professional misconduct to tailor training to real-world challenges.
    \item Support continuous education: Develop and promote robust continuing education programs for practicing engineers, centered on real-world case studies and emerging ethical challenges to maintain and elevate ethical awareness over the course of their careers.
    \item Engage in cultural change: Recognize that fostering ethical engineering practice requires more than enforcement; it involves actively cultivating a professional culture that prioritizes integrity, transparency, and public accountability at every level of practice.
\end{itemize}   
Together, these recommendations aim to bridge the gap between ethical theory and professional practice, ensuring that future and current engineers are well-equipped to uphold the highest standards of integrity in service to society.''

\section{Closing Thoughts}
Professional orders, such as the Ordre des ingénieurs du Québec (OIQ), play a vital role in promoting ethical stewardship in the engineering profession. As Van de Poel, et al. \citep{van2023ethics} emphasize, engineers have a deontological duty to prioritize public welfare. This duty is symbolized by the Iron Ring tradition, which serves as a reminder of the engineer's responsibility to society \citep{petroski2017iron}. Contrary to the notion of technological neutrality, technology reflects human values and systems, rather than being neutral \citep{van2015values}. This understanding highlights the importance of ethical considerations in engineering practice. The consequences of neglecting these considerations can be severe, as illustrated by the Quebec Bridge collapse \citep{petroski2017iron} and the SNC-Lavalin corruption scandal \citep{yeager2021political}.The OIQ plays a crucial role in promoting ethical engineering practices through its legally binding Code of Ethics (OIQ, 2023). This code outlines the professional obligations and responsibilities of engineers, ensuring that they prioritize public welfare and adhere to the highest ethical standards. The findings from our thematic analysis of professional misconduct cases in the Ordre des ingénieurs du Québec (OIQ)'s disciplinary register have significant implications for engineering education and ethics training. These results underscore the critical need for a comprehensive overhaul of how we approach engineering education, particularly in the realm of professional ethics and real-world practice.\\

Our research calls for a greater emphasis on continuing education and professional development for practicing engineers. Universities and professional bodies should collaborate to develop robust, case-study-based ethics training programs that address the specific types of misconduct identified in this study.\\

In conclusion, the complex landscape of professional misconduct revealed by our analysis demands a paradigm shift in engineering education. By grounding ethics and professional practice education in real-world examples and scenarios, we can better prepare future engineers to navigate the ethical challenges of their profession, ultimately enhancing public safety, trust, and the integrity of the engineering field.

\bibliography{custom}

\end{document}